\documentstyle[multicol,aps,epsfig]{revtex}
\tighten
\draft
\begin{document}
%
%
\title{Compression of finite size polymer brushes} \author{T.A.
  Vilgis\footnote{Permanent address: Max-Planck-Institute for Polymer
    Research, Ackermannweg 10, 55128 Mainz, Germany}, A. Johner and J.-F.
  Joanny\\[2mm]}
\address{Institut Charles Sadron (UPR CNRS 022)\\
  6 rue Boussingault, 67083 Strasbourg Cedex, France\\
  L.E.A MPI Mainz/ICS Strasbourg} \maketitle
%
%
%
\bigskip
%
\begin{abstract}
We consider edge effects in grafted polymer layers under compression.
For a semi-infinite brush, the penetration depth of edge effects
$\xi\propto h_0(h_0/h)^{1/2}$ is
larger than the natural height $h_0$ and the actual height $h$.
For a brush of finite  lateral size $S$ (width of a stripe or radius of a
disk), the  lateral extension
$u_S$ of the border chains follows the scaling law $u_S = \xi 
\varphi (S/\xi)$. The scaling
function $\varphi (x)$ is
estimated  within the framework of a local Flory theory for  stripe-shaped
grafting
surfaces. For small $x$, $\varphi (x)$ decays as a power law in agreement with
simple arguments. The effective line tension and the variation with
compression height
of the force applied
on the brush are also
calculated.
\end{abstract}
%
%
%
%
%
\begin{multicols}{2}
\section{Introduction}
Grafted polymer layers are widely used to prevent colloidal aggregation, and
therefore stabilize colloidal solutions. They are also utilized
to modify the mechanical properties of solid - solid contacts\cite{napper}.
They have attracted considerable
theoretical attention\cite{halperin,milnerrev} since the pioneering work by
Alexander and de Gennes \cite{ADG,PGGb}.
The overall  thickness of the layer results from a balance between the
stretching energy of the
chains  and the  excluded volume interaction as shown in fig. (1). 
\begin{center}
\begin{minipage}{8cm}
\label{fig1}
\begin{figure}
\centerline{{\epsfig{file=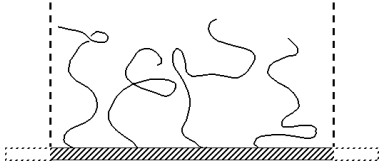,width=5cm}}}
\vspace{0.3cm}
\caption{{ 
A portion of an infinite brush : The chains stretch away from the
grafting surface to
lower their excluded volume interaction.}}
\end{figure}
\end{minipage}
\end{center}
\vspace*{.3cm}

Most of the work, however,  has been devoted to infinite
brushes. In this case a Flory theory or a blob argument can be easily
formulated to calculate the properties of the brush, i.e., its equilibrium
height and its free energy. In a Flory-type approach, the natural
thickness $h_0$ of a brush composed of linear chains of contour length $N$
at grafting density
$\sigma$ minimizes the free energy per unit  area that can be written as:
\begin{equation}
F = k_{\rm B}T \left(\sigma
h^2/Na^2 +
{v\over2}(\sigma N/h)^2h\right)
\;,
\end{equation}
where the first term stands for the
elastic energy and the second
for the excluded volume energy. In a good solvent, the
Edwards  excluded volume parameter $v$ is positive. The minimization with
respect to the height $h$
leads to the equilibrium height $h_{0}$ and the equilibrium free energy $F_{0}$
 \begin{equation}
h_0^3 = vN^3\sigma a^2/4 \quad \mbox{and} \quad F_0 /k_{\rm B}T = 3\sigma
N \left({\sigma v\over 4a}\right)^{2/3}
\; .
\label{natural}
\end{equation}
When the brush is compressed down to a height
$h<h_0$, the excluded volume term dominates over the elastic term. The
Flory-type free energy per unit area
$F_\infty$ can then be rewritten in order to emphasize the ratio
of the excluded volume to the  elastic energy. It is convenient for the
rest of the paper to introduce a correlation length $\xi$ and to write
the free energy of the brush with infinite extension (no edge effects)
\begin{equation}
F_\infty = k_{\rm B}T{\sigma h^2\over Na^2} \left(1 + {\xi^2\over 4h^2}\right)
\; .
\end{equation}
The length $\xi$ defined by $\xi^2 = 8 h_0^3/h$, is larger than both the
actual height
$h$ and the natural height $h_0$. Although the definition of $\xi$ might
not be obvious at
present, it will turn
out below that it plays a major role in the description of edge effects,
due to the fact that
the chains have more degrees of freedom close to the border of the brush
than in  the center.

Earlier, Rapha\"el and de Gennes proposed a description of the edge of a
unconstrained semi-infinite
brush  that captures the essential physics \cite{raphael}. They  describe
the edge effects on the free energy of the layer by a polarization
$P$ proportional to  the extension of the
chains parallel to the grafting plane. The elastic energy contribution due
to splay is then $\propto P^2$ and the next relevant
term in a Landau expansion accounting
for non uniform splay  is therefore
$\propto (\nabla P)^2$. The problem contains the unique length scale $h_0$ and
the splay at the edge
entails a negative effective line tension $\tau \sim - F_0h_0$.

A more refined description of polymer brushes originating in an idea by
Semenov \cite{sashablocks} was
proposed and developed over the last decade \cite{MWC,milnerrev}. This takes
advantage of the
drastic reduction of chain path fluctuations due to chain tension. Taking
into account only most
probable paths the model is exactly solvable, it gives access to the free
end density and predicts
the concentration profile. When the chain ends are constrained to the very
edge of the layer, a
situation similar to the Alexander model is recovered.

Several models have also been proposed in the
case where the chain paths are not parallel straight
lines\cite{fredrickson,xi}. A major difficulty
arises when the minimal paths are curved : it is then obvious that there is
not a unique minimal path
through a given point. To overcome this difficulty, either the free end
points have
been constrained (fixed) on the outer surface of the layer following
Frederickson and coworkers
\cite{fredrickson} or a
unique path description is used
in a locally parabolic molecular potential scheme\cite{xi}. Both models can
be solved for weakly
disturbed brushes. Milner's model seems very successful to estimate  free
energies as illustrated by
the calculation of block copolymer phase diagrams. Those models can be
looked at as variational
approaches. As we will be  concerned with finite to huge deviations from
the unconstrained brush we
propose here a simpler model where the chain trajectories are straight
lines and the free energy is
calculated in a local Flory theory. In a sense this is a crude version of
the previous models and
in line with the scaling arguments used later on. The close link to
the Rapha\"el - de Gennes
approach will also appear clearly.

The paper is organized as follows.
We first introduce and
describe our simple model. It is then
solved in the case of a semi-infinite
brush. In section III we
consider finite brushes before discussing the relevance of the present
work to
some experiments.

\section{The Model and the semi-infinite brush case}

In this section we introduce the local Flory model and compute the edge
effects in a semi - infinite brush. This is a grafted polymer layer which
has only one edge, whereas in the other direction the grafting surface is
infinite.
This case is useful to find the appropriate length scales.
Throughout the paper
we consider the chain trajectories to be straight lines. A chain
starting at the
grafting surface fills up a volume delimited by the nearby chains and the
confining surface.
The free energy is then a functional of the splay.

To be more specific, let us consider a
semi-infinite brush grafted on the half-plane $x<0$ with a uniform grafting
density $\sigma$ and
let the (large) length along the $y$-axis be $L_y$. A chain starting at x,
ends on the opposite
surface at $X(x)$  shows a splay $u(x) = X-x$ and fills a box of volume
$h\sigma^{-1}(1+u'/2)$,
with $u'=du/dx$ as shown in Figure 2. 
\begin{center}
\begin{minipage}{8cm}
\label{fig2}
\begin{figure}
\centerline{{\epsfig{file=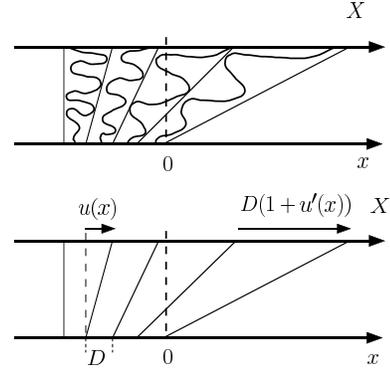,width=5cm}}}
\vspace{0.3cm}
\caption{{ 
The box model : A chain path is pictured as a straight line
defined by the grafting
coordinate $x$ and the splay $\rm{u}(x)$, the chain occupies a volume
defined by the nearby chains.}}
\end{figure}
\end{minipage}
\end{center}
\vspace*{.3cm}
The free energy functional reads
\begin{eqnarray}
F/k_{\rm B}T&=&\frac{L_y}{Na^{2}}\sigma \int\limits_{-\infty}^{0}dx\left[
u\left(
x\right) ^{2}+h^{2} \right]+\nonumber
\\
 &{}&v  L_y\frac{N^{2}\sigma ^{2}}{h}\int\limits_{-\infty}^{0}dx
\frac{1}{2+u'}
\; .
\label{energyfunctional}
\end{eqnarray}
The Euler Lagrange equation obtained after functional minimization of
eq. (\ref{energyfunctional}) with respect to the splay $u(x)$ has the
first integral,
\begin{equation}
u^2 + \xi^2 {1 + u'\over (2+u')^2} = C
\; ,
\label{firstintegral}
\end{equation}
where the characteristic length scale $\xi$ has been introduced previously.
Note that $\xi$ is much larger
than both $h$ and $h_0$ at high compression. The integration  constant $C$
is imposed by the boundary conditions. In the case of the
semi-infinite brush, where the splay $u$ and the deformation $u'$
vanish at $x=-\infty$, the appropriate choice is $C=1/4\xi^{2}$.
It is useful to introduce the dimensionless variables
 $\tilde{u}
= u/\xi$ and $\tilde x =x/\xi$. The
integration of eq.(\ref{firstintegral}) shows that the splay tail decays
exponentially over a characteristic length $\xi/4$,
\begin{equation}
\log{\tilde u\over \tilde u_0} - 2 (\tilde u - \tilde u_0) = 4 \tilde x
\label{splayeq}
\; .
\end{equation}
The splay $u_0$ towards  the very
edge remains to be determined. This is done by calculating the effective
line tension (i.e., the (negative) free energy cost of the splay per unit
length of the
border line) and
minimizing\footnote {For finite $L$, a case considered shortly, there is
indeed a minimum and a
maximum,
the latter being unphysical, here both extrema degenerate in an inflexion
point} it with respect to
$\tilde u_0$. We obtain
$\tilde u_0 = 1/2$ and the effective line tension :
\begin{equation}
\tau = - {k_{\rm B}T\over 48}{\sigma\over Na^2}\xi^3
\label{tau}
\; .
\end{equation}
As expected the line tension is negative as the splay relaxes  some  pressure
at the brush edge. In fact, $\tau$ represents a fraction of the
energy stored in the edge of size
$\sim\xi$.
Each chain in the edge contributes a in-plane tension
$\tau_{\rm i}\sim k_{\rm B}T \xi/N a^2$
and at the level of
scaling arguments, the in-plane tensions compensate  the osmotic
pressure inside
the brush.

The penetration depth $\xi$ has  a
simple physical meaning. The elastic energy of a
chain at the edge $U \sim \xi^2/a^{2}N$ compensates the excluded volume
energy of a bulk
chain. Alternatively, the chains
over the penetration length $\xi$ build up a brush  almost parallel to the
grafting surface with an
effective grafting density
$\sigma_{\parallel}\sim\sigma\xi/h$  with a brush "height" $\xi$,
as shown in Figure 3.
\begin{center}
\begin{minipage}{8cm}
\label{fig3}
\begin{figure}
\centerline{{\epsfig{file=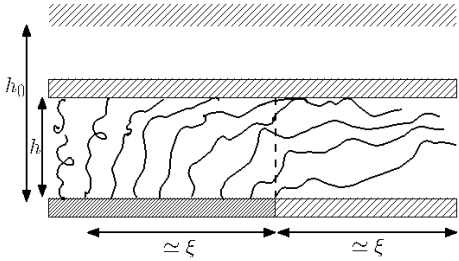,width=6cm}}}
\vspace{0.3cm}
\caption{{ The penetration depth $\xi$ of the edge effect : The chains
located within  the
penetration depth build a brush parallel to the grafting surface at density
$\sim
\sigma\xi/h$ per lateral unit area that extends over a width $\xi$.
}}
\end{figure}
\end{minipage}
\end{center}
\vspace*{.3cm}

When
calculating the excluded volume free energy in eq.(\ref{energyfunctional}),
we have assumed a continuous distribution of chains i.e.
$\xi/D>>1$, with
$D=\sigma^{-1/2}$ the distance between neighboring grafted chains, which is
fair. For $du/dx<<1$ we
could further expand the
excluded volume energy, the first relevant term is then
$\propto(\nabla u)^2$ similar to the
square gradient term of Rapha\"el and de Gennes mentioned above \cite{raphael}.
The final minimization of the line tension with respect to the splay at the
edge introduces a
weakly singular solution at the border (the derivative
$du/dz$ diverges there in order to
ensure a vanishing
concentration) that is however of no serious consequence. The present
very simple model forces the boxes to be filled up with monomers.
This assumption is reasonable as long as the polymer "pushes" on
the surfaces but  obviously fails at the very edge where the chain boxes
are very large and we nevertheless assume that the chain stretch over
the largest box size. The model also allows for
half of the available space only to be occupied. At the level of scaling
arguments our approach is
sound, however we cannot expect very accurate prefactors or scaling
functions. A better model allows
the chain to choose how much of the box height it wants to occupy. Let $H$ be
the box height occupied
by the chain, the volume occupied by the chain becomes $ H\sigma^{-1}
(1+u'H/ 2h)$, $H$ minimizes the
chain chemical potential, this links $H$ to the splay derivative:
\begin{equation}
\left({H\over 4h_0}\right)^3 = {1 +u'H/ h\over (2+u'H/ h)^2 (1 + u^2/h^2)}
\label{notfilled}
\; .
\end{equation}
The simple model with filled boxes is valid as long as $H>h$ i.e. using
eq.(\ref{notfilled})
for values of the splay up to
$\xi/\sqrt 8$ for strongly compressed brushes and for a vanishing splay in an
uncompressed brush as
 expected (though $u_0\sim
h_0$ remains obviously qualitatively correct for the uncompressed brush).
In the remainder we use the simple box model in combination with
scaling laws.

As shown above,  chains over the penetration depth $\xi$ form a brush
parallel to
the grafting surface of "height" $\xi$. This merely states that there is a
unique length scale
governing the edge effects. Assuming this to remain true when the chains
follow the excluded volume
statistics where
$h_0\sim N\sigma^{1-\nu\over\nu (d-1)}$, we obtain the scaling behavior:
\begin{equation}
\xi\sim h_0 \left({h_0\over h}\right)^{1-\nu\over \nu d-1}
\end{equation}
with $\nu$ the swelling exponent and $d$ the space dimension. The splay
relaxes the constraint
over the  edge where a finite fraction of the free energy is saved, this
corresponds to an
effective line tension:
\begin{equation}
\tau\sim  \tau_0\left(h_0\over h\right)^{2-\nu\over d\nu-1}
\end{equation}
where $\tau_0\sim -F_0h_0$ is the line tension for the uncompressed brush and
$F_0\sim \sigma N
\sigma^{1\over \nu(d-1)}$.

\section{Finite Size Effects}

In a semi-infinite brush, edge effects penetrate over the distance $\xi\sim
h_0(h_0/h)^{1/2}$; finite size effects thus become important when the
lateral extension
of the brush becomes
comparable to $\xi$. To be more specific, let us take a stripe of width $2L$,
extending between $x=-L$
and $x=+L$, uniformly grafted with a density $\sigma$. The simple box model can
again be solved exactly.
The functional minimization of the free energy with respect to the splay $u(x)$
leads to the same
Euler-Lagrange equation as in the semi-infinite case and the first integral
given by
eq.(\ref{firstintegral}) still holds. However, the integration constant $C$
is no  longer
$1/4\xi^2$.  The integration constants are fixed by the requirements of
vanishing
splay at the stripe center $u(0)
= 0$ and of minimal line tension. After integration of
eq.(\ref{firstintegral}), the splay is then
found to obey:
\begin{eqnarray}
4\tilde x &=& -2 \tilde u + \mbox{arcsinh} (\tilde u \delta)\nonumber\\
\mbox{where:}\quad \delta &=& {\tilde u_L\over\mbox{sinh}\left(4 L/\xi + 2
\tilde
u_L\right)}\nonumber\\
\mbox{with:} \quad  {2L\over\xi} &=& {1\over 2}\mbox{arctanh} 2\tilde u_L
-\tilde u_L 
\label{splaystripe}
\end{eqnarray}
For $L\gg\xi$ the semi-infinite brush result is recovered. At low values of
$L/\xi$ however the reduced splay at the border vanishes as a power law
$\tilde u_L = (3 L/2\xi)^{1/3}$.
Thus in the strongly compressed bush, half of the chains of the
stripe extend on either
side in a brush of height $\sim N(\sigma L/h)^{1/3}$, as shown in Figure 4.
\begin{center}
\begin{minipage}{8cm}
\label{fig4}
\begin{figure}
\centerline{{\epsfig{file=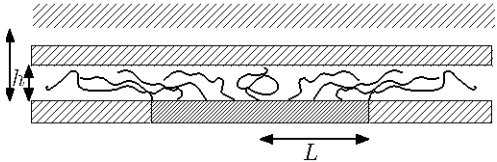,width=6.5cm}}}
\vspace{0.3cm}
\caption{{  Finite size effects : the grafted stripe. At high compression the
penetration depth is
bound by the half-size $L$. On either side, half of the chains build a
brush parallel to the grafting
surface at density $\sigma L/h$.
}}
\end{figure}
\end{minipage}
\end{center}
\vspace*{.3cm}
The scaling form for
$u(L)=u_L$ is given by :
\begin{eqnarray}
u(L)& =& \xi {\varphi}(L/\xi)\quad   \\
&\mbox{with:}&\quad
\lim_{x\rightarrow\infty}{\varphi}(x) = 1/2 \nonumber \\
&\mbox{and}&
{\varphi}(x)\sim (3x/2)^{1/3}\quad (x\ll1) \nonumber
\; .
\end{eqnarray}
\begin{center}
\begin{minipage}{8cm}
\label{fig5}
\begin{figure}
\centerline{{\epsfig{file=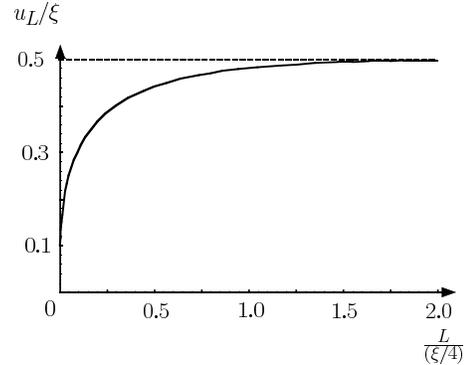,width=6cm}}}
\vspace{0.3cm}
\caption{{ Scaling of the splay at the border (stripe) : the reduced splay
at the border $u_L/\xi$ is
plotted against the reduced half-width $L/(\xi/4)$ (we choose $\xi/4$ as
unit width because it is the
decay length of the splay).
}}
\end{figure}
\end{minipage}
\end{center}
The complete  scaling function, as obtained from eq.(\ref{splaystripe}) is
plotted in Figure 5. Assuming excluded volume statistics we obtain instead
$\varphi (x)\sim x^{1-\nu\over\nu
(d-1)}$ for $x\ll1$.

During compression, the layer evolves from a  brush with chains oriented
perpendicular to
the grafting surface to a brush with chains oriented almost parallel to the
surface. This entails a
qualitative change in the force. For $L/\xi\ll1$, the free energy is that of a
brush parallel to the surface
$\sim Nh(\sigma L/h)^{5/3}$ and accordingly $f\sim N(\sigma
L/h)^{5/3}$.

The explicit free energy can be computed from eq.(\ref{splaystripe})
\begin{equation}
F = k_{\rm B}T{\sigma L L_y\over Na^2} \left(- {u_L^3\over 6L} +
{\xi^2\over 4} +
h^2\right)
\; ,
\label{energystripe}
\end{equation}
and shows how the "edge" term (first term) cancels against the osmotic term
in the absence of
splay (second term) to leading order at high compression.
After derivation with respect to the brush height $h$ we obtain the force,
\begin{equation}
f = k_{\rm B}T{LL_y\sigma\xi^2\over 4 Na^2h}\left({3\over 2}  - { \tilde
u_L^3\over  L/\xi} - 2 \tilde
u_L^2 -8{h^2\over\xi^2}\right)
\; .
\label{stripeforce}
\end{equation}
At large width, the edge effects only contribute  a
$\xi/L$ edge correction. For small width ($L/\xi\ll1$), the force scales as
$f \sim f_{\rm osm}(L/\xi)^{2/3}$; here
$f_{\rm osm} \simeq k_{\rm B}T{LL_y\sigma\xi^2\over 4 Na^2h}$ is the
osmotic force
in the absence
of any edge
effect. The pressure on the layer increases as
$L^{2/3}$.
Pressure versus distance profiles are shown in Figure 6.
The pressure unit is
the osmotic
pressure in the absence of any splay.

If the chains have  excluded volume statistics, we can still consider
qualitatively that a
brush parallel to the grafting surface is formed at high compression. The
free energy per unit area
now scales as
$\sim
\sigma_{\rm eff} N \sigma_{\rm eff}^{1\over \nu(d-1)}$ 
with $\sigma_{eff}=\sigma
L/h$. The total free energy
takes the simple form
$F(L)\sim L L_y F_\infty (L/\xi)^{1\over
\nu(d-1)}$ $(L\ll\xi)$, the force just scales as $f\sim F(L)/h$.
\begin{center}
\begin{minipage}{8cm}
\label{fig6}
\begin{figure}
\centerline{{\epsfig{file=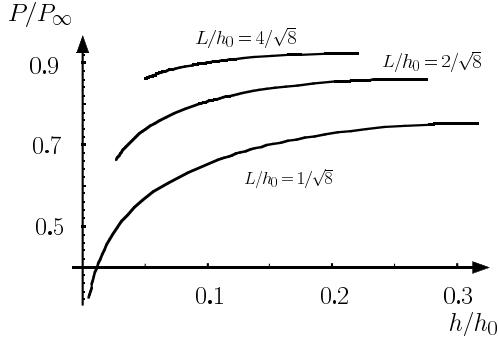,width=6.5cm}}}
\vspace{0.3cm}
\caption{{ Pressure versus height profiles (stripe). The pressure exerted to
confine the brush (force
divided by the grafted surface) as a function of the brush height.
$P_\infty$ is the osmotic
pressure exerted by an infinite brush of the same height (at the heights
considered the elastic
contribution is almost negligible), $h_0$ is the free brush height.
}}
\end{figure}
\end{minipage}
\end{center}
\vspace*{.3cm}

We now briefly consider a brush grafted on a finite disk of radius $R$,
Figure 7. For a large
disk ($R\gg\xi$) the curvature has no effect and the splay at the border
$u(R)$ is of order $\xi$. 
\begin{center}
\begin{minipage}{8cm}
\label{fig7}
\begin{figure}
\centerline{{\epsfig{file=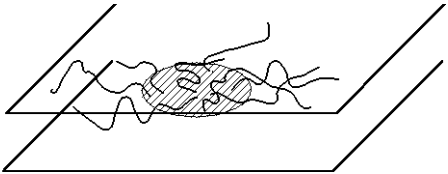,width=6.5cm}}}
\vspace{0.3cm}
\caption{{ The grafted disk. At high compression, the polymers escape
laterally and form a
cylindrical brush.
}}
\end{figure}
\end{minipage}
\end{center}
\vspace*{.3cm}
In
general we anticipate a scaling form: $u(R) = \xi \varphi (R/\xi)$. For
small disks, we expect the formation of a
cylindrical brush that involves all grafted polymers. It is easily
checked that most of the
monomers belong to the corona extending outside the grafted disk. The local
correlation length
$\xi_\phi$ at a distance $r$ from the center follows from a Daoud and
Cotton\cite{daoudcotton} type
of argument,
$\xi_\phi^2\sim {r h\over\sigma R^2}$. The monomer concentration at a
distance $r$
from the disk center is 
$c(r)\simeq \left( \frac{\sigma R^2}{ rh} \right)^{3\nu-1/(2\nu)}$.
The conservation of monomers then leads to:
\begin{equation}
u(R)\sim \xi\left({R\over\xi}\right)^{2(1-\nu)\over 1+\nu}\quad (R/\xi\ll1)
\end{equation}
This holds at not too high compressions, as long as $\xi_\phi(R)<h$
otherwise a dilute corona is
build beyond $r_{\rm dil}$ given by $\xi_\phi (r_{\rm dil}) = h$
where the polymers
are constrained by the
plates rather than by the other polymers\cite{halpjoanny}. In the regime
where the polymers are
semi-dilute through the whole corona, the free energy and the force obey
the simple scaling form :
\begin{equation}
F(R)\sim F_\infty R^2\left(R\over\xi\right)^{2\over 1+\nu}\quad f(R)\sim F(R)/h
\label{forcedisk}
\end{equation}
and the pressure increases faster with the radius of the disk than with the
width of the stripe,
mainly because the polymer can escape in two dimensions for a disk
and only in one
dimension for a stripe (note that a formal increase in the dimension
of space does not add
any escape dimension for the polymer but only increases the longitudinal
dimension of the stripe).

\section{Conclusion}

We have presented a study of the compression of finite size and
semi-infinite brushes based on a
Flory-type of calculation and scaling arguments. Chains close to the border
relax some of
their excluded volume interaction by creating a marked outward splay, which
decreases towards
the center
of the brush.

For semi-infinite brushes, the very tail of the splay profile decays
exponentially over a
length $\sim \xi$. The penetration depth of the edge effects $\xi$ is much
larger
than both the natural
brush height
$h_0$ and the imposed height $h$; it is given by
$\xi\sim h_0\left(h_0/h\right)^{1-\nu\over 3\nu-1}$, and it is found to be the
only relevant length
scale. This means that chains grafted within the penetration depth $\xi$
form a brush almost
parallel to the grafting surface of extension $\sim\xi$; this in turn is
sufficient to determine the
length $\xi$. The relaxation of the confinement constraint at the edge
entails an
effective negative line tension $\tau\sim -F_\infty \xi$ where $F_\infty$
is the free energy
per unit area
in an infinite brush.

Finite size effects occur when the lateral size $S$ of the layer (width L
of a grafted stripe, radius
R of a grafted disk) becomes of order $\xi$. The splay $u_S$ at the border
is  weaker than in the
semi-infinite case : $u_S\sim \xi\varphi (S/\xi)$ with $\varphi (x)$ of
order one for large
argument and $\varphi (x)\propto x^\alpha$ for small arguments. The
exponent $\alpha$ can be
obtained from simple arguments: at high compression, all chains are
collected to build a brush
parallel to the grafting surfaces. The value obtained for a  stripe $\alpha
= {1-\nu\over 2\nu}$
is slightly smaller than that obtained for a disk $\alpha = {2(1-\nu)\over
1+\nu}$.

As the brush (at least partially) escapes laterally during compression, the
pressure $P$
exerted to confine the layer (evaluated with respect to the fixed grafted
surface) increases with
the brush size. More precisely, $P\sim P_\infty
\left(S\over\xi\right)^\beta$
where the value $\beta = 1/(2\nu)$ for a stripe is smaller than that for a
disk $\beta =
2/(1+\nu)$.

The layer is softer against compression
with increasing border
length: for a given grafted area, this means that layers divided in many
small  pieces,
collections of stripes, or
layers with
 holes are softer than compact layers. If the border lines are curved,
positive curvatures (with the
center of curvature located
on the layer side) also make the brush softer. R\"uhe and coworkers
synthesize grafted
layers with controlled
in plane structures. They are able to obtain long grafted chains by
$\em{in}$ $\em{situ}$ radicalar
polymerization on surfaces where lines seeded with (or free of) initiators
are drawn via
interference patterns. Those systems should be excellent model systems to
study the mechanical
response to compression.

The (weaker) negative effective line tension for unconstrained brushes has
been used by Rapha\"el and
de Gennes\cite{raphael} to explain the formation of ribbon-shaped or
disk-shaped aggregates in
flexible(soluble)/rigid(insoluble)/flexible(soluble) triblock copolymers.
One would expect the larger
line tension in constrained systems to increase the length of the border
line by formation
of holes or cutlines.

 When a  diblock copolymer
lamella (or stack of lamellas) is compressed, if the compression
is released the lamellar period has to increase back to its equilibrium
value. The characteristic
time for
an overall relaxation diverges with some (almost) macroscopic size. We thus
expect relaxation  to occur
via the formation of defects (cuts and holes). A microscopic model such as the
one developed here is
needed to study the formation of these defects and their subsequent healing.

A related topic is the formation of stable defects close to a
phase boundary in
amphiphile/water systems\cite{roux}. A model very similar to the one
presented here
based on polymer physics could
provide some molecular
expressions for phenomenological quantities used in surfactant physics.

Microgels may spread on an adsorbing surface at the expense of elastic energy
\cite{stapper}. 
This can be seen as a compromise between surface tension and confinement.
Confinement
of the microgel starting from the swollen $c^*$ state can be studied in a
way very similar to the study of
polymer brushes presented here.
To be specific take a piece of microgel of constant thickness irreversibly
pasted on a flat
solid by one face and
adsorbing onto a parallel substrate by the opposite face.
A very naive model for the gel is a perfect cubic gel that deforms upon
confinement
as the brush with the extra chain segments (crosslinking the brush chains)
remaining parallel
to the substrate. A length scale $\xi$ for the penetration of edge effects can
again be defined as $\xi\sim h_0(h_0/h)^{1/2}$ where $h_0$ is the initial
swelling equilibrium
height. At the level of scaling arguments, balancing surface tension and
confinement
energy leads to the spreading condition (contact surface much larger than
grafting
surface), $\gamma > \sigma^{4/3} L^2/h_0$ in the case of a stripe
(using mean-field exponents). The same  spreading condition is obtained for the
brush, however it is not clear whether the simple model used
in this paper to describe the edge is suitable in this case.

\section*{acknowledgement}
TAV greatfully acknowledges the kind hospitality of the Insitut Charles Sadron (CRM-CNRS) and the Laboratoire Europ\'een Associ\'e for financal support.

\end{multicols}

\end{document}